\documentclass[draftclsnofoot,12pt,onecolumn]{IEEEtran}

\usepackage{moreverb}
\usepackage{graphicx}
\usepackage{multirow}
\usepackage{amsmath}
\usepackage{pxfonts}
\usepackage{txfonts}
\usepackage{graphicx}
\usepackage{multirow}
\usepackage{mathrsfs}
\usepackage{eso-pic}
\usepackage{fancyhdr}
\usepackage{amsmath}
\usepackage{algorithm}
\usepackage{algpseudocode}
\usepackage{graphicx}
\usepackage{tabularx}

\DeclareMathOperator{\st}{subject\ to}

\begin{document}

\title{Low complexity resource allocation for load minimization in OFDMA wireless networks}

\author{Ying Yang$^1$,Marco Moretti$^2$,  Wenxiang Dong$^1$, Weidong Wang$^1$\\
$^1$University of Science and Technology of China Hefei, China\\(e-mail: yying@mail.ustc.edu.cn)\\
$^2$Information Engineering Department - Universit\`a di Pisa, Via Caruso 16, 50126 Pisa, Italy (e-mail: marco.moretti@iet.unipi.it)
}
\maketitle

\begin{abstract}
To cope with the ever increasing demand for bandwidth, future wireless networks will be designed with reuse distance equal to one. This scenario requires the implementation of techniques able to manage the strong multiple access interference each cell generates towards its neighbor cells. In particular, low complexity and reduced feedback are important requirements for practical algorithms. In this paper we study an allocation problem for OFDMA networks formulated with the objective of minimizing the load of each cell in the system subject to the constraint that each  user meets its target rate. We decompose resource allocation into two sub-problems: channel allocation under deterministic power assignment and continuous power assignment optimization. Channel allocation is formulated as the problem of finding the maximum weighted independent set (MWIS) in graph theory. In addition, we propose a minimal weighted-degree greedy (MWDG) algorithm of which the approximation factor is analyzed.  For power allocation, an iterative power reassignment algorithm (DPRA) is proposed. The  control information requested  to perform the allocation is limited and the computational burden is shared between the base station and the user equipments. Simulations have been carried out under constant bit rate traffic model and the results have been compared with other allocation schemes of similar complexity. MWDG has excellent performance and outperforms all other techniques.
\end{abstract}


\section{Introduction}
\label{secintro}
Providing broadband wireless access with a guaranteed quality of service (QoS) to an ever increasing number of users  is one of the major challenges for future wireless communication systems.  Orthogonal frequency division multiplexing (OFDM), due to its robustness to multipath fading and low-complexity  implementation, is the transmission technology adopted for future mobile communication systems \cite{Chuang&Sollenberger2000}. Orthogonal frequency division multiple access (OFDMA), the multiple access scheme based on OFDM technology, partitions the available bandwidth in orthogonal channels for the users in a cell and as such allows the deployment of channel-aware dynamic  radio resource allocation  to fully  exploit frequency and multiuser diversities \cite{MoPN13}. Due to the scarcity of spectrum, over the years, research has been focused on increasing systems' spectral efficiency  and this has lead to the design of full  frequency reuse networks \cite{Boudreau2009}, where all cells transmit on all available resource blocks simultaneously. In this scenario,  strong   inter-cell interference (ICI),  represents the most limiting  factor to the performance of future mobile communication systems.
Several schemes have been researched to address and solve the problem of ICI so to  improve the performance of OFDMA systems as discussed in \cite{Moretti2011},\cite{Ksairi2010} and the references therein. In general, ICI is addressed by  solving complex optimization problems. The main limit of this approach is the computational complexity and the need of exchanging large amounts of information: to efficiently allocate radio resources the base station (BS) may need to know the channel gains and the interference levels measured on all available resource blocks.
As an alternative approach, research has focused on inter-cell interference coordination (ICIC) \cite{Liang2012}: techniques designed to minimize the interference experienced in the network and maximizing spatial reuse. Among ICIC schemes, fractional frequency reuse (FFR) has received a lot of attention, \cite{Xu2012}, \cite{Yu2013}. The basic concept of FFR is to employ a small frequency reuse factor for cell-center users who have higher signal-to-interference-plus-noise ratios (SINRs) and a large one for cell-edge users who have lower SINRs. In this way, FFR can achieve good tradeoff between the average throughput of the entire network and the performance of the cell-edge users \cite{Chang2009}.

Mainstream resource allocation has solved the problem of either minimizing the transmitted  power under target rate constraints \cite{Letaief&Murch1999}\nocite{kim2006}-\cite{Pischella08} or maximizing the overall rate under a power constraint \cite{Gong2011}-\nocite{Yu06}\cite{Yang2011}. In this paper, rather than focusing on one of the two previous approaches,  the allocation problem is formulated with the objective of minimizing the load of each cell in the system, i.e., the number of occupied physical resource blocks per cell, with the constraint that each served user meets its target rate request. This formulation, amenable to a distributed per cell implementation, is particularly useful in a multi-cell scenario, where the load of a cell directly influences the amount of interference in neighboring cells. Moreover, by minimizing the load of each cell, the processing complexity or power consumption  can be reduced as well\cite{powerload}. In case it is not possible to satisfy all users' request, the proposed algorithm aims at maximizing the number of served users enforcing a load control policy \cite{Gong2011}.
To reduce the allocation complexity, the original allocation  problem has been divided into two sub-problems: channel assignment  under deterministic power distribution and, subsequently, power allocation. With the help of graph theory, channel assignment is formulated as a maximum weighted independent set (MWIS) problem and it is solved by means of a heuristic technique for whose performance we find an approximation guarantee. Power allocation is an iterative strategy designed to further reduce the cell load.

The practical feasibility of the proposed scheme is guaranteed by its low complexity and the little amount of control traffic required. The computational load of the allocation is split between the user terminals and the base stations. Each user signals to the base station (BS) the set containing all the possible combinations of radio resources that  fulfils its target rate so that the  BS is able to solve the MWIS problem by means of a simple heuristic. Other algorithms \cite{Yang2011}, \cite{Abr12} are designed to share the computational load of the allocation between user terminals and the BS but in general they require the exchange of large amounts of information on dedicated control links while in this case the information exchanged is restricted to just the set of channels that fulfils a user's target rate.
\par
The remainder of this paper is organized as follows. Section \ref{secsysmo} introduces the system model and the allocation problem we studied. Section \ref{secPRBallo} proposes a suboptimal low-complexity  formulation of the allocation problem, amenable to a distributed per cell implementation. In Section \ref{subsectiongraph} we present a  greedy heuristic, based on graph theory, for the  solution of the problems described in the previous section. In Section \ref{secDPRA} an iterative distributed power reassignment algorithm (DPRA) is studied for further load minimization. The implementation of the proposed algorithms is discussed in Section \ref{secimpdis}. Section \ref{secperf} presents the numerical results and compare the performance of the proposed scheme with other algorithms. Finally, conclusions are drawn in Section \ref{secCon}.
\par
\emph{Notation}: Sets, matrices and vectors are denoted by boldface letters, the notation $|\mathcal{S}|$ indicates the cardinality of  set  $\mathcal{S}$.

\section{System model and problem formulation}
\label{secsysmo}
We consider a cellular system where  the whole available spectrum is shared among all cells, i.e., the frequency reuse is
equal to one. Let $\Omega$ be the set collecting all the cells, in each cell $i \in \Omega$ there are a BS and $S^{(i)}$ mobile users. The modulation technique is multi-carrier and the available spectrum is partitioned into a set of orthogonal subcarriers. Adjacent subcarriers are grouped in a physical resource block (PRB), which is the basic allocation element. Let $\pi$ be the set of all the PRBs so that the total bandwidth $F$ is spanned by the elements of $ \pi$ and the bandwidth of a single PRB is $B=F/\left|\pi\right|$.
\par
The BS schedules the users and assigns them a subset of the
radio channels. We formulate the allocation problem with the objective of minimizing the cell load, defined as the total number of used PRBs, with the constraints of transmitting a certain target rate $R_u^{(i)}$ for each user $u \in S^{(i)}$ and each cell $i \in \Omega$.  Focusing on a certain quality of service, those users that can not achieve the target rate will be dropped.
\par
By indicating with $p_n^{(i)}$ the power transmitted on sub-channel $n$ in cell $i$, and using the Shannon capacity as a measure of the transmitted rate,  the achievable transmission rate of  user $u$ served by BS $i$ on  PRB $n$ is:
\begin{equation}
\label{eqs1}
r_{u,n}^{(i)} = B{\log _2}\left(1 +\frac{p_{n}^{(i)} h_{u,n}^{(i,i)} }{ I_{u,n}^{(i)}+ \sigma^{2}}\right)
\end{equation}
where $I_{u,n}^{(i)}=\sum\limits_{j\in\Omega,j\ne i}p_{n}^{(j)} h_{u,n}^{(j,i)}$ is the multiple-access interference affecting user $u$ and $h_{u,n}^{(j,i)} $ is the squared gain on channel $n$ between the BS $j$ and user $u \in S^{(i)}$ and  $ \sigma^{2}$ is the thermal noise power. In the following we are considering a population of static or slow-moving users so that the propagation channel has a long coherence time.
\par

Let $\mathbf{X}_i$ ($i=1,\dots,\left|\Omega\right|$) be the  $\left| S^{(i)}\right|\times |\pi|$-dimensional allocation  matrix for cell $i$, where ${X_i}(u,n)=1$ if PRB $n$ is allocated to user $u$ and  0 otherwise. Let $\mathbf{P,X}$ be the vectors obtained by stacking all the power and channel allocations, the multi-cell load minimization allocation problem (MCLMAP) can be formulated as
\begin{align}
\min\limits_{\mathbf{P,X}} \quad &\sum\limits_{i \in \Omega} \sum\limits_{n \in \pi } \sum\limits_{u \in S^{(i)}}X_i(u,n)   \label{eqs3}\\
\st \quad &
\sum\limits_{u\in S^{(i)}} {{X_i}(u,n)}  \le 1\quad \forall i \in \Omega,  \forall n \in \pi \tag{\ref{eqs3}.1} \label{C1}\nonumber \\
& \sum\limits_{n \in \pi } r_{u,n}^{(i)}X_{i}(u,n)  \ge R_{u}^{(i)} \quad \forall i \in \Omega, \forall u \in S^{(i)} \tag{\ref{eqs3}.2}\label{C2}\nonumber\\
&\sum\limits_{n \in \pi } {p_i^n}  \le P^{(i)} \quad \forall i \in \Omega \tag{\ref{eqs3}.3}\label{C3}\nonumber\\
&\sum\limits_n X_i(u,n)  \le M\quad \forall i \in \Omega, \forall u \in S^{(i)} \tag{\ref{eqs3}.4} \label{C4}\nonumber \\
& X_i(u,n) \in \left\{ {0,1} \right\} \quad \forall i \in \Omega, \forall n \in \pi, \forall u \in S^{(i)} \tag{\ref{eqs3}.5}\label{C5} \nonumber
\end{align}

The set of constraints \eqref{C1} indicates that each PRB can be assigned to  at most one user per cell, the set of constraints \eqref{C2} dictates the rate requirement for all the users in the system, the set of constraints \eqref{C3} makes sure that the overall power transmitted in cell $i$ does not exceed the total power  $P^{(i)}$, and the set of constraints \eqref{C4}  limits the maximum number of PRBs per user to $M$ which is introduced  to prevent a hungry user with a bad link from  occupying a large number of PRBs and making the system too unfair.  Finally, the constraints \eqref{C5} account for the fact that the allocation variable can only take the values 0 and 1.
\par
MCLMAP belongs to the class of  mix-integer programming (MIP): the simultaneous presence of continuous (power) and binary (allocation indicator)  variables makes this type of problems very hard to solve. The interference term in the rate computation in \eqref{C2} further complicates its solution. Moreover, a centralized formulation such as \eqref{eqs3} requires the knowledge of the channel gains for all users on all PRB in all cells determining a large exchange of control information. Therefore,    we propose a heuristic solution for MCLMAP, amenable to a distributed per-cell implementation, based on solving power and allocation optimization in two separate steps.

\section{Single cell resource allocation allocation under deterministic power assignment}
\label{secPRBallo}

To simplify the solution of MCLMAP, we propose a suboptimal low-complexity approach based on  dividing it into two separate subproblems: PRB allocation and power allocation.

To perform resource allocation, we initially distribute uniformly the power on all PRBs in all cells, so that it is $p_{n}^{(i)}=P^{(i)}/\left|\pi\right|$ ($\forall i \in \Omega, \forall n \in \pi$). By fixing the power assignment matrix $\mathbf{P}$, one determines also the interference between the cells  as any PRB allocation will not change the interference pattern. Under these assumptions, interference is  decoupled from PRB allocation and each BS  can  allocate its PRBs independently. Since the rate on a given channel for a given user is a constant, the load minimization can now be solved as an integer programming problem.   Without loss of generality, we focus on cell $i \in\Omega$  to formulate the single-cell fixed-power load minimization allocation problem (SCFPLMAP) as
\begin{align}
\min\limits_{\mathbf{X}_i} \quad &  \sum\limits_{n \in \pi } {\sum\limits_{u \in S^{(i)}} {{X_i}(u,n)} }  \label{eq2-1} \\
 \st \quad & \sum\limits_{u \in S^{(i)}} {{X_i}(u,n)}  \le 1\quad \forall n \in \pi \tag{\ref{eq2-1}.1} \label{C2.1}\nonumber \\
&\sum\limits_n r_{u,n}^{(i)}X_i(u,n)  \ge R_{u}^{(i)}\quad \forall u \in S^{(i)} \tag{\ref{eq2-1}.2} \label{C2.2}\nonumber \\
&\sum\limits_n X_i(u,n)  \le M\quad \forall u \in S^{(i)} \tag{\ref{eq2-1}.3} \label{C2.4}\nonumber \\
&{X_i}(u,n) \in \left\{ {0,1} \right\}\quad  \forall u \in S^{(i)}, \forall n \in \pi  \tag{\ref{eq2-1}.4} \label{C2.3}\nonumber
\end{align}

Given certain channel configurations and users' rate requirements, it may happen that not all the users are able to meet their target rates and SCFPLMAP can not find a feasible solution. In this case,  the allocation problem must be modified and reformulated with the objective of serving the largest possible number of satisfied users. Let $\mathcal{S}^{(i)} = \left\{u|u \in S^{(i)},\sum\limits_{n\in \pi}  r_{u,n}^{(i)}X_i(u,n)  \ge R_{u}^{(i)},\sum\limits_n X_i(u,n)  \le M\right\}$ be the set containing all satisfied users in cell $i$, our goal is to  maximize the number of elements of $\mathcal{S}^{(i)}$. In this scenario,  allocation is formulated as a single-cell fixed-power  admission control problem  (SCFPACP)
\begin{align}
\max\limits_{\mathbf{X}_i} \quad &  \left|\mathcal{S}^{(i)}\right|   \label{eq2-1-1} \\
 \st \quad & \sum\limits_{u \in S^{(i)}} {{X_i}(u,n)}  \le 1\quad \forall n \in \pi \tag{\ref{eq2-1-1}.1} \label{C2.1.1}\nonumber \\
&\sum\limits_n X_i(u,n)  \le M\quad \forall u \in S^{(i)} \tag{\ref{eq2-1-1}.3} \label{C2.1.3}\nonumber \\
&{X_i}(u,n) \in \left\{ {0,1} \right\}\quad  \forall u \in S^{(i)}, \forall n \in \pi  \tag{\ref{eq2-1-1}.2} \label{C2.1.2}\nonumber
\end{align}

\section{A graph model for SCFPLMAP and SCFPACP}
\label{subsectiongraph}

In this  section we present an equivalent graph model for both SCFPLMAP and SCFPACP. By describing the allocation problems with a graph, we can show that they  can be both modeled as max-weighted-independent-set (MWIS) problems. Although MWIS problems are NP-hard, we propose a low-complexity heuristic to solve them in linear time.

For user $u \in S^{(i)}$, the \emph{minimal resource allocation set} $Y_u=\left\{ {{Y_{u,1}},...,{Y_{u,j}},....} \right\}$ is defined as the set containing all the subsets $Y_{u,j} \subseteq  \pi$ such that a) $\sum\limits_{n\in Y_{u,j}}  r_{u,n}^{(i)}  \ge R_{u}^{(i)}$ and b) $Y_{u,{j_1}} \not\subset Y_{u,{j_2}} \hspace{3pt}\forall j_1,j_2$.  Condition a) guarantees that the resources in  $Y_{u,j}$ allow user $u$ to meet its target rate and condition b) requires that any element in ${Y_u}$  contains the minimum possible number of PRBs  to satisfy the user rate requirements.
\par
To proceed further we need to introduce the concepts of clique and independent set in graph theory.
\par
{\bf{Definition 1}:} \emph{Given an undirected graph $G = (V,E,W)$  a subset of nodes $S\subseteq V$ is a \emph{clique} if every pair of nodes in $S$ has an edge between them. A subset of nodes $S \subseteq V$ is an \emph{independent} set if there is no edge in $E$ between any two nodes in $S$. Given a graph $G = (V,E,W)$ the \emph{max-weighted-independent-set} (MIWS) is the independent set in $G$ that has maximum weight.}
\par
The equivalent graph $G = \left( {V,E,W} \right)$, designed to describe the allocation problems SCFPLMAP and SCFPACP,  is built from the minimal resource allocation sets of all the users in $S^{(i)}$ by applying the following rules:
\begin{itemize}
\item Every vertex $v$ of $G$ represents a set $Y_{u,j}$ in ${Y_u}$ and  its weight is  computed as the total number of available PRBs minus the number of PRBs in $Y_{u,j}$, i.e.,  $W(v) = |\pi | - |Y_{u,j}|$.
\item  All the vertices corresponding to the  sets in ${Y_u}$ are connected so that  ${Y_u}$ is represented by a clique in $G$;
\item If  $Y_{u_1,j_1}\in Y_{u_1}$ and $Y_{u_2,j_2}\in Y_{u_2}$ share at least a PRB, then an edge connects  the corresponding vertices of $G$.
\end{itemize}
\par
 By construction, taking into account that the maximum number of resources per user is constrained by \eqref{C2.4} to $M$ PRBs,  the weight of each vertex of $G$ is bounded by $ |\pi |-M\leq W(v)\leq  |\pi |-1$.
 \par
After having constructed the graph  $G$ describing SCFPLMAP and SCFPACP allocation problems, we can formulate the following theorem.
\par
{\bf{Theorem 1}:}  \emph{SCFPLMAP and SCFPACP can be modeled as  MWIS problems.}
\begin{IEEEproof}
We first focus on SCFPLMAP. In cell $i$ if SCFPLMAP is feasible, then all users $u\in S^{(i)}$ meet their target rate and we indicate with $\mathcal{Y}_{1}$ the independent set corresponding to the SCFPLMAP allocation. Let us denote with $v_{u}^{\ast}$ the node in the graph $G$ corresponding to the PRB allocation for user $u$ in the solution of SCFPLMAP and, stretching  a bit the notation, with $\left|v_{u}^{\ast} \right|$  the  number of PRBs associated with the vertex $v_{u}^{\ast}$, then the weight of $\mathcal{Y}_{1}$ can be lower bounded as $W\left(\mathcal{Y}_{1}\right)=\sum\limits_{u\in S^{(i)}}W(v_{u}^{\ast})=\left|S^{(i)}\right|\left|\pi \right|-\sum\limits_{u\in S^{(i)}}\left|v_{u}^{\ast} \right|\ge\left|S^{(i)}\right|\left|\pi \right|-\left|\pi \right|=\left(\left|S^{(i)}\right|-1 \right)\left|\pi \right|$, where the inequality follows from $\sum\limits_{u\in S^{(i)}}\left|v_{u}^{\ast} \right|\le \left|\pi \right|$.
\par
\emph{By contradiction}, suppose that SCFPLMAP is not a MWIS problem and it exists an independent set $\mathcal{Y}_{2} \ne \mathcal{Y}_{1}$ that has maximum weight over $G$, i.e., $W\left(\mathcal{Y}_{2}\right)>W\left(\mathcal{Y}_{1}\right)$. This proof is organized in two parts depending if the  the set $\mathcal{Y}_{2}$ contains or not all the subsets associated to all users in cell $i$.
\par
If $\mathcal{Y}_{2}$ does not include all users, then it contains at most the sets corresponding to $\left|S^{(i)}\right|-1$ users and therefore its weight can be upper bounded as $W\left(\mathcal{Y}_{2}\right)\le\left(\left|S^{(i)}\right|-1\right)\left(\left|\pi \right|-1\right)$.  Comparing this bound with the bound found for $\mathcal{Y}_{1}$  shows that $W\left(\mathcal{Y}_{1}\right)>W\left(\mathcal{Y}_{2}\right)$, but this contradicts the hypothesis.
\par
If $\mathcal{Y}_{2}$ does contain all the subsets associated to the users in cell $i$, let us denote with $v_{u}^{\left(\mathcal{Y}_{2}\right)}$ the vertex in $\mathcal{Y}_{2}$ corresponding to the PRB allocation for user $u$, so that it is $W\left(\mathcal{Y}_{2}\right)=\left|S^{(i)}\right|\left|\pi \right|-\sum\limits_{u\in S^{(i)}}\left|v_{u}^{\left(\mathcal{Y}_{2}\right)} \right|$. Since $\mathcal{Y}_{1}$ represents the SCFPLMAP solution it is $\sum\limits_{u\in S^{(i)}}\left|v_{u}^{\ast} \right|\le \sum\limits_{u\in S^{(i)}}\left|v_{u}^{\left(\mathcal{Y}_{2}\right)} \right|$ and $W\left(\mathcal{Y}_{2}\right)\le W\left(\mathcal{Y}_{1}\right)$, which contradicts the hypothesis, hence SCFPLMAP is a MWIS problem.

If SCFPLMAP is unfeasible, some users need to be dropped and the allocation problem is formulated as SCFPACP.  By contradiction, we need to show that it exists an independent set in $G$ that has a weight greater than the solution found by SCFPACP while serving a smaller number of users. Following the first part of the proof adopted for SCFPLMAP is straightforward to show that SCFPACP is a MWIS problem.
\end{IEEEproof}

 Unfortunately, when $M>1$ the MWIS problem is  \emph{NP-hard}  \cite{Robson1986}-\nocite{Fomin2009}\cite{Karp2010}. Instead, if we set $M=1$,  each  vertex' weight is $ |\pi |-1$ and the MWIS problem reduces to the max-independent-set (MIS) problem. Since each vertex contains one PRB only, each vertex and its neighbor vertexes form a clique in the graph and the optimal independent-set can be obtained in polynomial time with greedy algorithms.

\subsection{Minimal Weighted Degree Greedy Algorithm}

Since MWIS problems are NP-hard, in this subsection we propose a greedy heuristic designed to solve with low complexity SCFPLMAP and SCFPACP as  MWIS problems.
\par
By construction, each minimal resource allocation set is translated into a clique in $G = \left( {V,E,W} \right)$ so that the graph is composed by $\left|S^{(i)}\right|$ cliques. Let  $\mathcal{C}_u \subset V$ with $u\in S^{(i)}$ represent  the clique of vertices spawned by user $u$. The quality of the allocation associated to the vertex $v \in \mathcal{C}_u$ can be measured by two parameters: the \emph{weight} and the \emph{degree}, defined as the number of edges incident to $v$. A large weight is  attractive since it indicates that the specific allocation requires a low number of PRBs. On the contrary, a large degree indicates either that there are many vertices in the clique $\mathcal{C}_u$ and user $u$ has many other allocation choices or that the specific  allocation associated to $v$ requires a set of PRBs that are requested by many other users. 
\par
Let $\mathcal{P}_{k}(v) \subseteq \mathcal{C}_{k}$ 
be the set of vertices in $\mathcal{C}_{k}$ adjacent to $v$, the \emph{weighted degree} of the vertex $v \in \mathcal{C}_u$ is defined as
\begin{equation}
\label{eq2--1}
{d_w}\left( v \right) = \frac{\sum\limits_{v_\ell \in \mathcal{C}_{u},v_{\ell}\ne v} W\left( v_\ell\right)+\sum\limits_{k \in S^{(i)}/\left\{u\right\}} \sum\limits_{v_j \in \mathcal{P}_{k}(v)} W\left( v_j\right)}{W(v)}.
\end{equation}

The definition of the weighted degree is similar to that in \cite{Kako2005} which is expressed as the ratio of the sum-weight of adjacent vertices to the weight of itself. A vertex with a low weighted degree has either a large weight or it is adjacent to  a small number of vertices and in both cases represents a 'good' allocation. Accordingly, the \emph{ minimal weighted degree greedy} (MWDG) algorithm is a heuristic designed with the objective of  iteratively selecting the vertex with minimum weighted degree  among all the vertices of the graph $G$. The  MWDG algorithm can be summarized as:

\subsubsection{Initialization}
\begin{quote}
\begin{itemize}
\item Build the graph  $G = \left(V,E,W\right)$.
\item Initialize $V^{(1)}=V$, $E^{(1)}=E$ and the allocation matrix $\mathbf{X}_{i}=\mathbf{0}$.
\item Compute the weighted degree ${d_w}\left( v \right)$ for all   $v \in V^{(0)}$.
\end{itemize}
\end{quote}

\subsubsection{Iteration}
\begin{quote}
During the $j^{th}$ iteration the  MWDG algorithm performs these operations
\begin{itemize}
\item Select in $ V^{(j)}$ the vertex  with the minimal weighted degree:
\begin{equation}
\label{eq2-4}
m^{(j)} = \arg {\min _{m \in V^{(j)} }}{d_w}\left( m \right)
\end{equation}

\item Update the  allocation matrix $\mathbf{X}_{i}$. Suppose that vertex $m^{(j)}$ is associated with the allocation set $Y_{u,k}$ of user $u$, then the $u$th row of the allocation matrix is updated by setting $X_{i}(u,n)=1$, $\forall n\in Y_{u,k}$.
\item Update the vertex set $V^{(j+1)}=V^{(j)}-\mathcal{M}^{(j)}$, where  $\mathcal{M}^{(j)}\subseteq V^{(j)} $ is the set that contains the vertex $m^{(j)}$ and all vertices in $V^{(j)}$ adjacent to  $m^{(j)}$.
\item Update the edge set $E^{(j+1)}$ by removing all the edges incident with the vertices in $\mathcal{M}^{(j)}$.
\item Update  the weighted degree ${d_w}\left( v \right)$ for all   $v \in V^{(j+1)}$.
\end{itemize}
\end{quote}

The MWDG algorithm is iterated until there are no more vertices in $V^{(j+1)}$, i.e. $V^{(j+1)}=\emptyset$.
\par
\begin{figure}[!htp]
   \centering
    \includegraphics[width=3in]{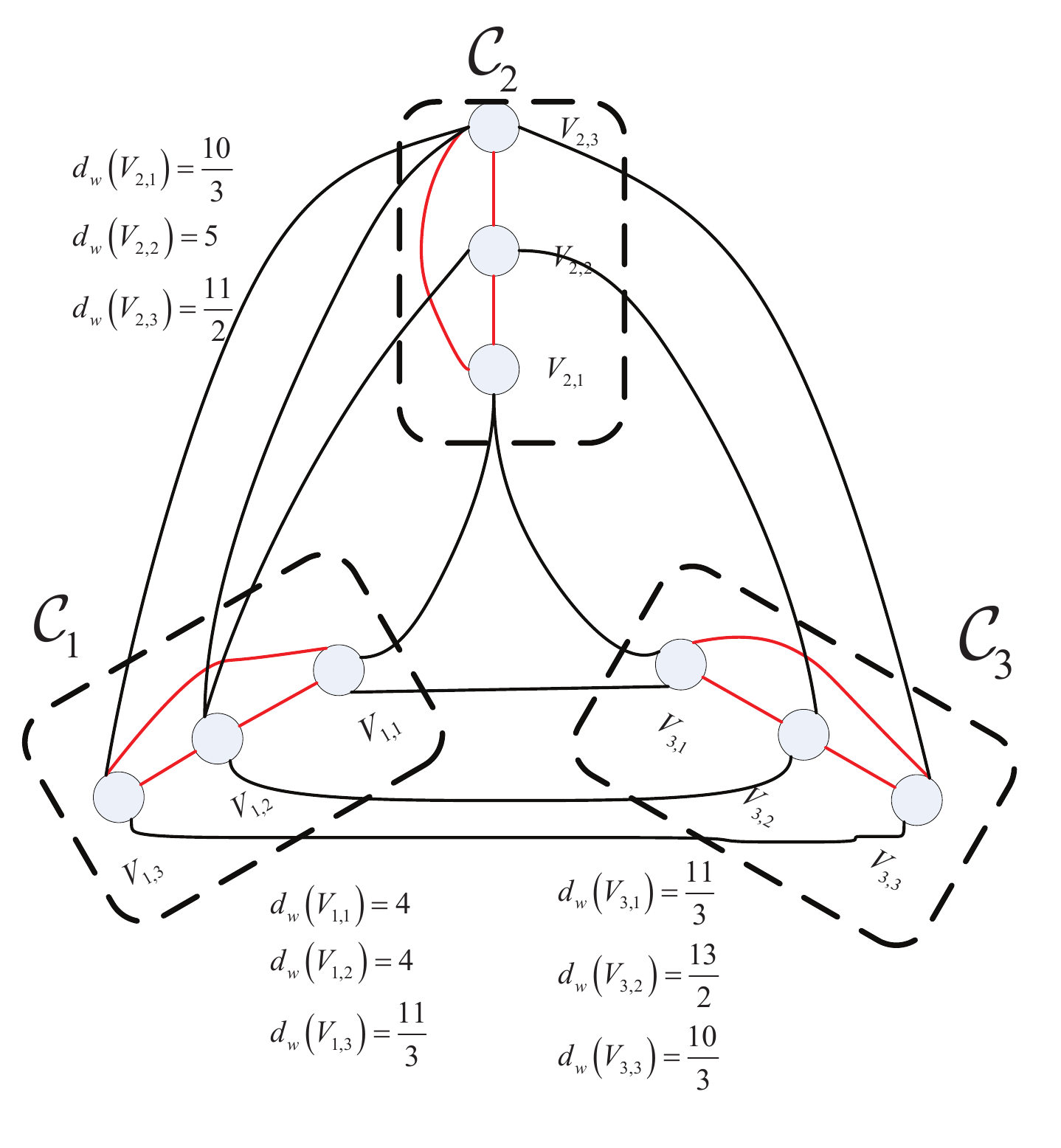}
     \caption{A toy example with three users and four PRBs.}
     \label{fig:Simple}
\end{figure}

Fig. \ref{fig:Simple} shows the graph $G = \left(V,E,W\right)$ for a toy example with three users and four PRBs. Suppose the users set is $S^{(i)}=\{u_1,u_2,u_3\}$, the PRBs set is $\pi=\{m_1,m_2,m_3,m_4\}$ and $M=2$. Furthermore, we assume that the minimal resource allocation sets are: $Y_{u_1}=\{\{m_1\},\{m_2\},\{m_4\}\}$, $Y_{u_2}=\{\{m_1\},\{m_2,m_3\},\{m_2,m_4\}\}$, $Y_{u_3}=\{\{m_1\},\{m_2,m_3\},\{m_4\}\}$.

\begin{figure}[!htp]
    \centering
    \includegraphics[width=3in]{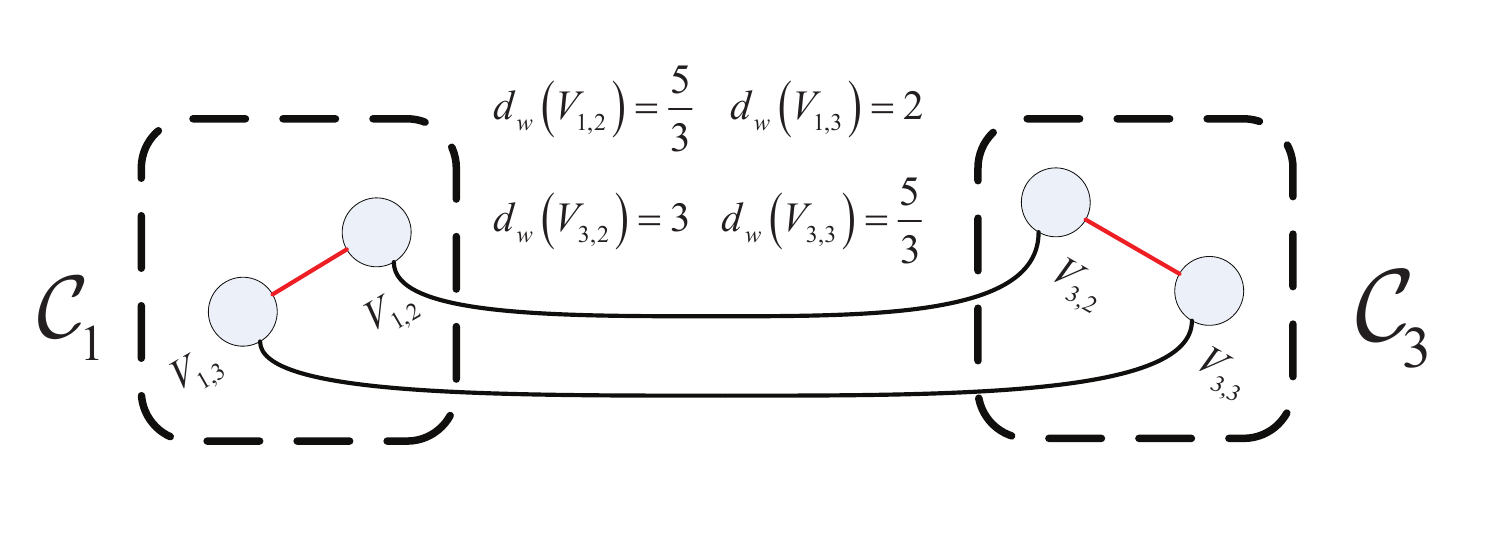}
     \caption{The graph after the first iteration of MWDG.}
     \label{fig:Simple2}
\end{figure}

The vertices that minimize the weighted-degree are $V_{3,3}$ and $V_{2,1}$ and any of the two can be  chosen indifferently. Assume MWDG selects vertex $V_{2,1}$ so that the PRB set $\{m_1\}$ is allocated to user $u_2$. As a consequence,  vertex $V_{2,1}$ and all  adjacent  vertices are removed from the graph and the resulting graph is shown in  Fig. \ref{fig:Simple2}, where the weighted-degree of the remaining vertices has been re-calculated and vertices $V_{1,2}$ and $V_{3,3}$ will be selected.

\subsection{Approximation Guarantee of the MWDG Algorithm}
\label{subsectionapp}
The \emph{approximation ratio} $\rho$ bounds the weight of any MWDG solution as a fraction of the weight of the optimal solution.

\par
{\bf{Theorem 2}:} \emph{Given the graph $G(V,E,W)$, let $W_{MWIS}\left( G \right)$ be the weight of the MWIS and $W_{MWDG}\left( G \right)$ be the weight of the independent-set obtained by the MWDG algorithm. The \emph{approximation ratio} $\rho$ of the MWDG algorithm is}
\begin{equation}
\rho=\mathop {\max }\limits_{G} \frac{W_{MWIS}\left( G \right)}{W_{MWDG}\left( G \right)}={M}\max\left(\frac{\left(|\pi|-2\right)}{|\pi|-M},1\right)
\end{equation}
\begin{IEEEproof}
Let $J$ be the cardinality of the independent set found by means of MWDG algorithm on the graph $G(V,E,W)$. Let $m^{(j)}$ be the vertex found at iteration $j$ and  $\mathcal{M}^{(j)}$ is the set that contains the vertex $m^{(j)}$ and all vertices in $V^{(j)}$ adjacent to  $m^{(j)}$, by construction it is
\begin{equation}
\label{eqUnion}
V = \bigcup\limits_{j=1}^{J} \mathcal{M}^{(j)}
\end{equation}

Let $H^{(j)}$ be the induced graph on the vertex set $ \mathcal{M}^{(j)}$ $(j=1,\dots,J)$. By construction $m^{(j)}$ is adjacent to all vertices in $ \mathcal{M}^{(j)}$ and therefore the weight of the MWIS in $H^{(j)}$ is either $W\left(m^{(j)}\right)$ or the weight of MWIS on $\tilde{H}^{(j)}$, the graph induced on the set  $\tilde{\mathcal{M}}^{(j)}=\mathcal{M}^{(j)}/m^{(j)}$,  so that the following relationship holds
\begin{equation}
\label{eq2-9}
W_{MWIS}\left( H^{(j)} \right) = \max \left(W\left(m^{(j)}\right),W_{MWIS}\left( \tilde{H}^{(j)} \right) \right)
\end{equation}

\par
In Appendix A we show that $W_{MWIS}\left( \tilde{H}^{(j)}  \right)$ can be bounded as:
\begin{eqnarray}
\label{eq2-10}
W_{MWIS}\left( \tilde{H}^{(j)}\right) &\le& \left( |\pi| -  W\left(m^{(j)}\right)  \right)\left( {|\pi| - 2} \right) \nonumber\\
&\le& {M}\left( {|\pi|- 2} \right)\nonumber\\
&\le& W\left(m^{(j)}\right)\frac{{{M}\left( {|\pi| - 2} \right)}}{{\left( {|\pi| - {M}} \right)}}\;.
\end{eqnarray}

Substituting \eqref{eq2-10} in \eqref{eq2-9} yields
\begin{equation}
\label{eq2-11}
W_{MWIS}\left( H^{(j)} \right) \le \max \left( W\left(m^{(j)}\right),W\left(m^{(j)}\right)\frac{M\left( |\pi| - 2 \right)}{\left( |\pi| - M \right)} \right).
\end{equation}

Because of \eqref{eqUnion}, the weight of the MWIS can be bounded by the sum of the weights of the MWIS of graphs $H^{(j)}$ $(j=1,\dots,J)$, i.e.
\begin{eqnarray}
\label{eq2-12}
W_{MWIS} \left( G \right) &\le& \sum\limits_{j=1}^{J} W_{MWIS}\left( H^{(j)} \right) \nonumber\\
& \le& {M}\max\left(\frac{\left(|\pi|-2\right)}{|\pi|-M},1\right)\sum\limits_{j=1}^{J} W\left(m^{(j)}\right)\nonumber\\
  &=& {M}\max\left(\frac{\left(|\pi|-2\right)}{|\pi|-M},1\right)W_{MWDG}\left( G \right).
\end{eqnarray}

The proof is completed and the approximation ratio of MWDG algorithm is $\rho={M}\max\left(\frac{\left(|\pi|-2\right)}{|\pi|-M},1\right)$.
\end{IEEEproof}

\emph{Remark}: Let the number of satisfied users with the optimal MWIS and MWDG algorithms be $U_{MWIS}$ and $U_{MWDG}$, respectively, and let the number of allocated PRBs with MWIS and MWDG algorithms be $N_{MWIS}$ and $N_{MWDG}$. It follows that
\begin{equation}
\rho \geq \frac{U_{MWIS}|\pi|-N_{MWIS}}{U_{MWDG}|\pi|-N_{MWDG}}
\end{equation}

so that the minimum number of satisfied users obtained with the MWDG algorithm can be lower bounded as $U_{MWDG}>U_{MWIS}/\rho$.
Furthermore, when it is $M=1$, the approximation ratio of MWDG algorithm is $\rho=1$, which means  that the MWDG algorithm can get the optimal independent set. As $M$ increases, then the approximation ratio will increase as well, expanding the gap between the  MWDG solution and the  MWIS optimum. However,  the system can be dimensioned so that reasonable values for $M$ are $1\leq M\leq 3$ \cite{Kwan2010} and the loss with   respect to MWIS is limited.

\section{Distributed Power Reassignment Algorithm}
\label{secDPRA}
Once the MWDG algorithm has found the allocation matrix $\mathbf{X}_{i}$ solution of the problems \eqref{eq2-1} or \eqref{eq2-1-1}, the initial uniform power distribution can be modified to further enhance the system performance. Keeping in mind that the allocation objective is to optimize the system's load, this section proposes a heuristic low-complexity strategy designed to reduce the users' power consumption with the goal of curtailing the multiple access  interference so that it is possible to further lower the cell load. Following a different design approach with respect to other power distribution algorithms such as \emph{scale} \cite{Papandriopoulos2009} or \emph{iterative waterfilling} \cite{Yu2004}, the distributed power reassignment algorithm (DPRA) aims at \emph{reducing the load} rather than minimizing the overall consumed power.
\par
As it might happen that certain PRBs are not allocated to any user, no power should be transmitted on the subcarrier in the set $\pi_{\emptyset}^{(i)}=\left\{n|\sum\limits_{u\in S^{(i)}}X_{i}(u,n)=0\right\}$. Moreover, as a consequence of a fixed predetermined power distribution, the overall rate of user $u$ in cell $i$ might exceed its target $R_{u}^{(i)}$. Accordingly, the power allocated to user $u$ should be reduced with the objective of freeing some resources and lowering the cell load.

\par
The DPRA machinery is illustrated in Algorithm \ref{alg1}. First of all, DPRA sets the transmitted power to zero on the PRBs which are not used. Then, DPRA tries either to reduce the transmitted power or the number of occupied PRBs of those users whose rate exceeds their target. In details,  since all users are allocated a PRBs set of minimal dimension, at the first iteration the DPRA can only lower the power of each users on the subcarrier where the power saving is the largest. This in turn causes a reduction of the multiple access interference in all  cells, so that in subsequent iterations  it might be possible to decrease the number of allocated PRBs or, alternatively, further reduce the transmitted power.
\begin{algorithm}
\caption{DPRA Algorithm in cell $i \in \Omega$}
\label{alg1}
\begin{algorithmic}[1]
\State {Initialize:}
\State{Set to 0 the power of the PRBs that have not been allocated}
\State  {$p^{(i)}_{m}=0, \forall m \in \pi_{\emptyset}^{(i)}$ }
\State Repeat:
\For {user $u\in S^{(i)}$}
\State  Build $\pi_{u}^{(i)}=\left\{n|X_{i}(u,n)=1,u\in S^{(i)}\right\}$
\State {Compute  the  rate $\Delta r_{u}$ in excess}
\State { $\Delta r_{u}= \sum\limits_{n\in\pi_{u}^{{(i)}}}r^{(i)}_{u,n}-R_{u}^{(i)}$}
\State {When possible reduce the cell load}
\While {$ \min\limits_{n\in \pi_{u}^{(i)}}r^{(i)}_{u,n} \le \Delta  r_{u}$}
\State {$\tilde{m}=\arg\min\limits_{n\in \pi_{u}^{{(i)}}}r^{(i)}_{u,n}$}
\State {$\pi_u^{(i)}=\pi_u^{(i)}-\{\tilde{m}\}$}
\State {$p^{(i)}_{\tilde{m}}=0$}
\State {$\Delta r_{u}=\Delta r_{u}- r^{(i)}_{u,\tilde{m}}$}
\EndWhile
\For {PRB $n\in \pi_{u}^{(i)}$}
\State $\Delta p_{u,n} = 2^{r^{(i)}_{u,n}/B}\left(1-2^{-\Delta r_{u}/B}\right)\frac{I^{(i)}_{u,n}+\sigma^{2}}{h^{(i,i)}_{u,n}}$
\EndFor
\State {Reduce the power in excess}
\State {$m^{\ast}=\arg\max\limits_{n\in \pi_{u}^{(i)}}\Delta p^{(i)}_{u,n}$}
\State {$p^{(i)}_{m^{\ast}} = p^{(i)}_{m^{\ast}}-\Delta p^{(i)}_{u,m^{\ast}}$}
\EndFor
\State Each user feeds back the power assignment to the server cell.
\State Until allocation converges in all cells.
\end{algorithmic}
\end{algorithm}
%
Since at each iteration of DPRA the power transmitted in all cells on each PRB  either is smaller or does not change, the DPRA algorithm converges in few iterations.

\section{MWDG and DPRA Implementation}
\label{secimpdis}
For all allocation algorithms, the computational complexity and the amount of control traffic needed to implement the algorithm play a key role in assessing the algorithm's feasibility.  A very important characteristic of MWDG and DPRA is that most of the computations are performed locally so that the load of the allocation is shared between the BS and the terminals  and the amount of  information that the users  exchange with the BS is extremely limited.
\par
In details, to solve the SCFPLMAP and SCFPACP allocation, each BS distributes the power uniformly on the available channels. After measuring the achievable rate on all PRBs, each user $u$ computes its minimal resource allocation set ${Y_u}$, whose elements are the sets of PRBs that satisfy the user's rate constraint. All the  resource allocation sets  are computed locally and user $u$ signals to the BS only the elements of $Y_{u}$, amounting to a few bytes of control traffic. The advantage with respect to  other allocation algorithms is striking since most of them \cite{Moretti2011}, \cite{Pischella08} requires the explicit knowledge of channel gains and interference levels on all PRBs, which need to be exchanged on a dedicated control channel.  Having collected all users' reports, each BS executes the MWDG algorithm, whose complexity is linear in the number of edges and vertices \cite{Halldorsson1997}.
After having received the information regarding its PRB allocation, each user will perform locally the DPRA. All the measures on which the DPRA is based are taken by the user terminals and so are the DPRA decisions, so that at each iteration  the required power levels and the occupied PRBs are signaled back to the BS. Considering that the DPRA converges in a few iterations, the amount of control traffic  is also in this case very limited.

\section{Numerical results}
\label{secperf}
To evaluate the performance of the proposed MWDG  and  DPRA algorithms we consider a cellular system composed by $|\Omega|=7$ cells distributed on an hexagonal grid. Following an  uniform spatial distribution, an equal number $N$ of users is   generated in each cell, i.e.,  $S^{(i)}=N, \forall i \in \Omega$. Users are static or slow-moving  so that the propagation channel has a long coherence time. The  propagation channel is frequency selective with independent Rayleigh fading. The most important simulation parameters are summarized in Table \ref{table-2}.
\par
We compare MWDG with three other algorithms: a \emph{random greedy} (RG) allocation, where in each cell users are selected in random order and are allocated the best available PRBs until fulfillment of their constraints, a \emph{mean enhanced greedy} (MEG) allocation \cite{Nwamadi2011}, where the users are first sorted in ascending order on the base of their mean rate computed over all PRBs and then assigned the best available PRBs, and the \emph{fractional frequency reuse} allocation described as FFR-B in \cite{Chang2009}. In particular, the FFR-B algorithm, originally designed for best effort traffic,  is based on a fixed power distribution and on the construction of an interference graph on the base of which the PRBs are assigned to the users. All the three algorithms have been simulated using the same value of $M$, the maximum number of PRBs that can be assigned to a single user, employed with MWDG.  RG and MEG algorithms have the peculiarity that they need approximately the same amount of feedback required by MWDG, since the ordering of the best subcarriers can be performed at the  user terminal and signaled back at the BS. Our goal is a fair comparison between practical strategies, therefore all algorithms have comparable complexity. For this same reason we have avoided the comparison with other techniques that require  full knowledge of all the channel gains and have much larger computational complexity, such as the algorithms proposed in \cite{Moretti2011} and \cite{La2012}, for example.
\begin{table}[h]
  \centering
\caption{Simulation parameters}
  \begin{tabular}{||c|c||}
  \hline
        	System bandwidth& $F=5$ MHz ($|\pi|=24$ PRBs)\\
        \hline
         	Inter-cell distance & 500 m\\
        \hline
        Total power constraint & $P^{(i)}=43$ dBm $,\forall i \in \Omega$\\
        \hline
        Pathloss & $128.1+37.6\log_{10}(d)$\\
        \hline
        Shadowing fading & Log-normal (standard deviation 8 dB)\\
        \hline
	Traffic Model & Constant Bit Rate $R_{u}=768$ kbps\\
	\hline
	Max  number of PRBs per user  & $M=1,2,3$\\
	\hline
        \end{tabular}
\medskip
\label{table-2}
\end{table}
\par
Since the allocation algorithms are defined with the goal of minimizing the cell load (SCFPLMAP) and  maximizing the number of served users (SCFPACP), we employ the  number of dropped users per cell and the average number $\eta$ of PRBs per user  as performance indicators. If system is feasible, all users will be satisfied, and the minimization of load in equation \eqref{eqs3} is the same as the minimization of average number $\eta$ of PRBs allocated to each user. If system is unfeasible which means some users will be dropped, then maximization of the number of satisfied users also equals to the minimization of the $\eta$, since all PRBs will be allocated to users.

In general, letting $\mathcal{S}^{(i)}$ the set containing the satisfied users in cell $i$, $|S^{(i)}|-|\mathcal{S}^{(i)}|$ users are dropped because the allocation algorithm is not able to find a set of PRBs that fulfills their rate requirements. The  number $\eta$ of PRBs assigned to each user averaged on all the cells in the system is calculated as:
\begin{equation}
\label{eq14}
\eta=\frac{1}{\left|\Omega\right|}\sum\limits_{i \in \Omega}\frac{\sum\limits_{u \in \mathcal{ S}^{(i)}} |\pi_u^{(i)}|}{\left|\mathcal{S}^{(i)}\right|}.
\end{equation}

\begin{figure}[!htp]
    \centering
    \includegraphics[width=3in]{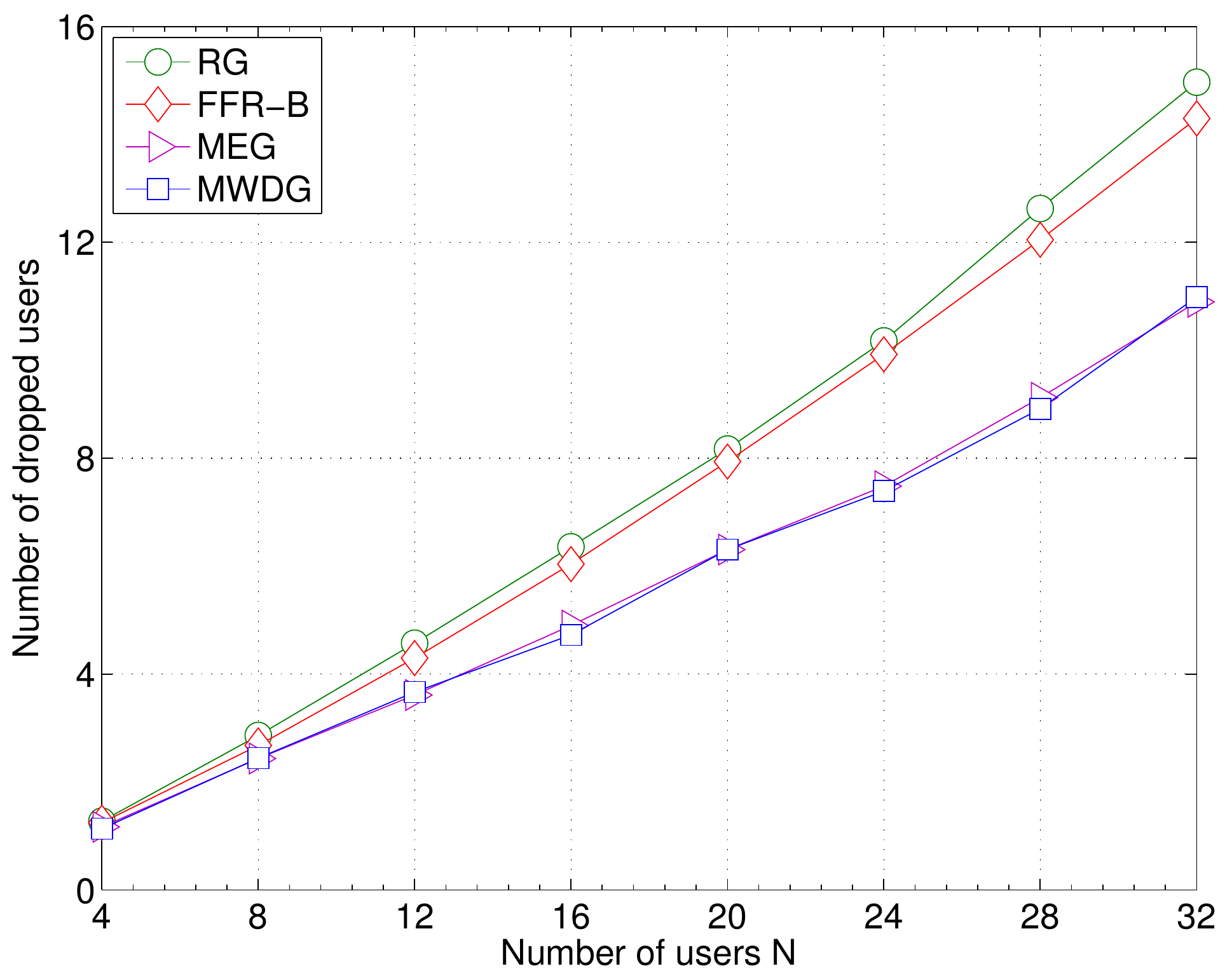}
     \caption{Number of dropped users vs.  number of users when $M=1$}
     \label{fig:dropk1}
\end{figure}

\begin{figure}[!htp]
    \centering
    \includegraphics[width=3in]{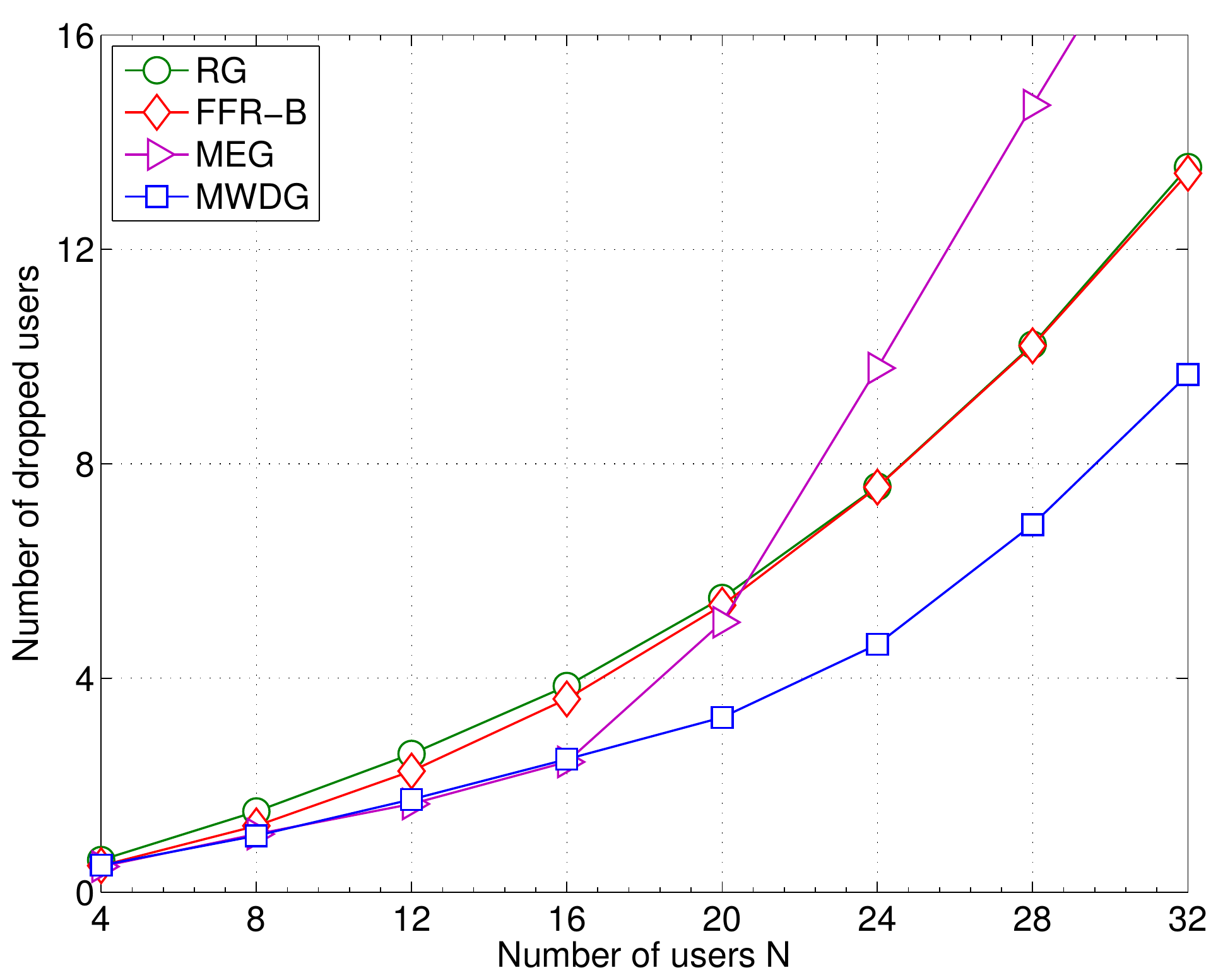}
     \caption{Number of dropped users vs.  number of users  when $M=2$}
     \label{fig:dropk2}
\end{figure}

\begin{figure}[!htp]
    \centering
    \includegraphics[width=3in]{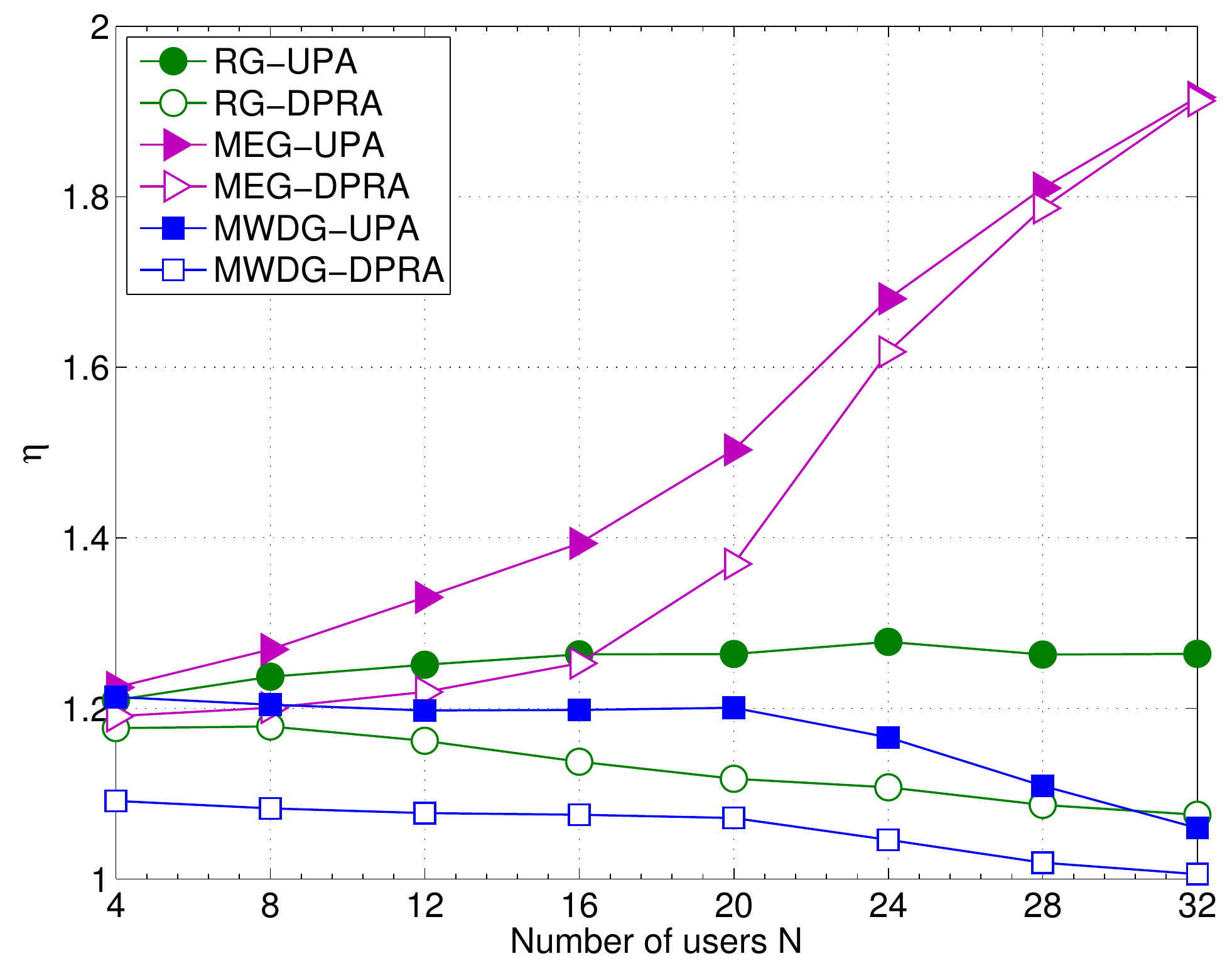}
     \caption{Average number of PRBs  per user $\eta$ vs. number of users when $M=2$}
     \label{fig:loadk2}
\end{figure}

\begin{figure}[!htp]
    \centering
    \includegraphics[width=3in]{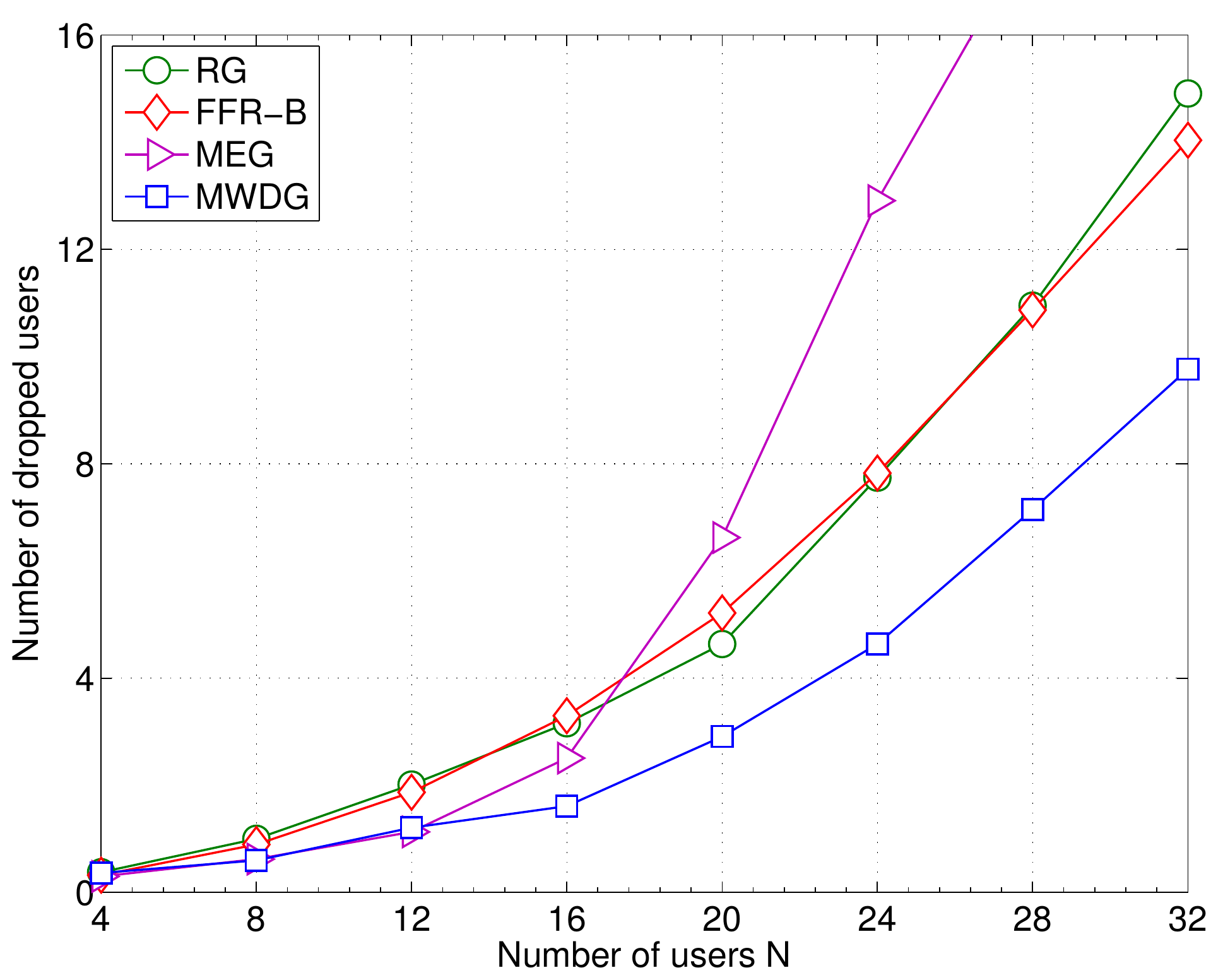}
     \caption{Number of dropped users vs.  number of users when $M=3$}
     \label{fig:dropk3}
\end{figure}

\begin{figure}[!htp]
    \centering
    \includegraphics[width=3in]{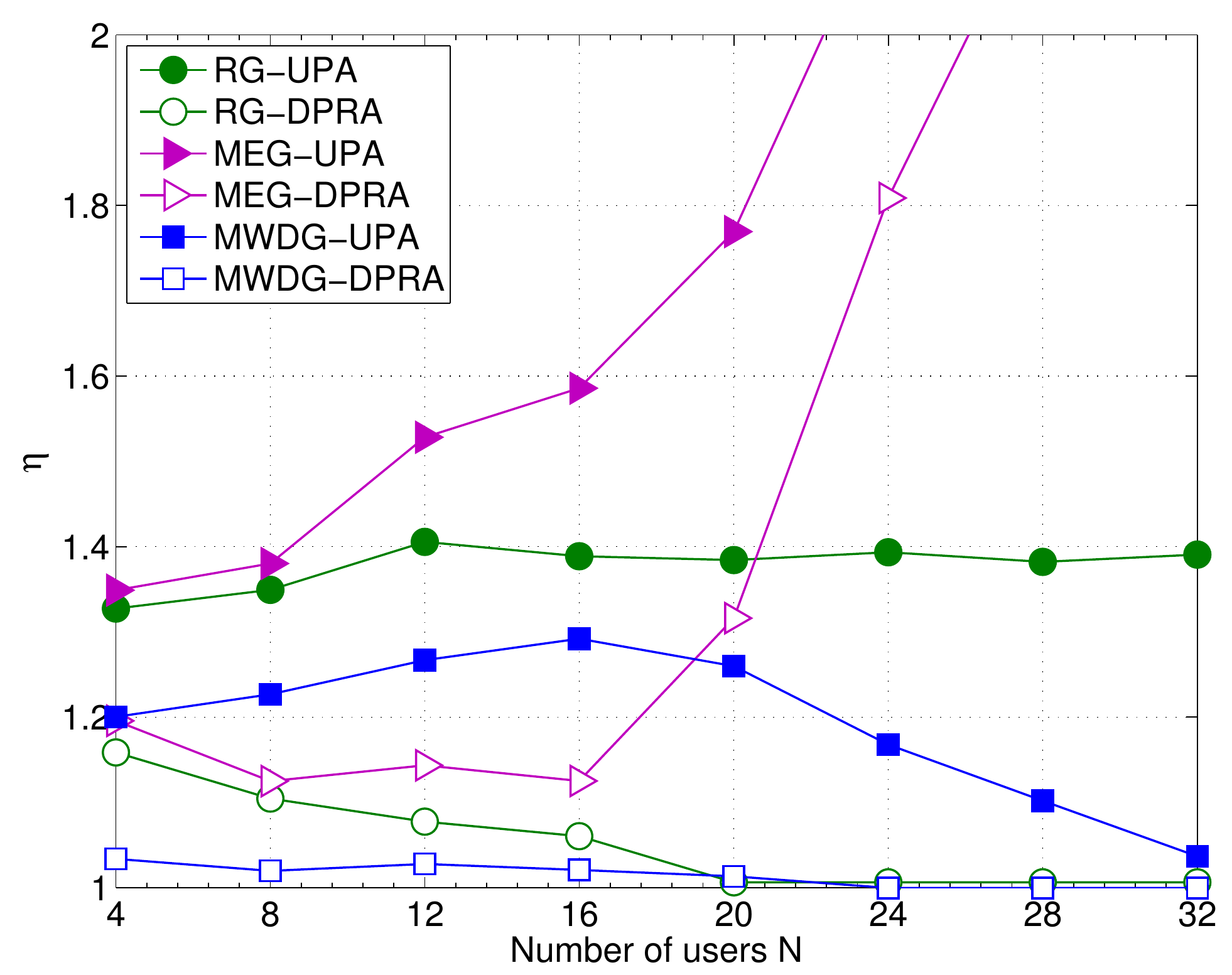}
     \caption{Average number of PRBs  per user $\eta$ vs. number of users when $M=3$}
     \label{fig:loadk3}
\end{figure}

Figs.~\ref{fig:dropk1}-\ref{fig:loadk3}  show the results for different values of $M$. The choice of $M$ plays a key role in assessing the allocator performance. On one hand, a large value of $M$ gives a lot of freedom to the allocator and allows users with bad propagation conditions to meet their requested rates. On the other hand,  the complexity of the MWDG algorithm is proportional to $M$ since the number of vertices of the allocation graph grows exponentially with it.
\par

Fig.~\ref{fig:dropk1}  shows the number of dropped users when it is $M=1$.  Results are plotted as a function of the number of users per cell. When the number of users is larger than the number of available PRBs, i.e.,  $N> |\pi|$, some users are necessarily dropped and MWDG solves the SCFPACP allocation. Nevertheless,  the number of dropped users is not negligible for small loads too. This is due to the fact that some cell edge users may not meet their requirements with just one PRB and they are necessarily dropped. The MWDG and MEG algorithms have very close performance: in this case,  the users with worst channels, which are the first to be served with MEG, have a small number of available PRBs sets and, accordingly, their vertices in the graph have a low  weighted degree  so that MWDG will serve them first  with high probability.  Since each user will be allocated only one PRB, then the average number of PRBs per user is  $\eta=1$ and the DPRA will not provide any performance gain in this scenario except for power  minimization. RG and FFR-B achieve similar results.
\par

Fig.~\ref{fig:dropk2}  plots the number of dropped users  for  $M=2$. In this case the number of dropped users for low cell loads is substantially reduced: most  users can meet their requirements with two PRBs. When the cell load is relatively small, i.e. $N \le 16$, MWDG and MEG drop approximately the same number of users. When the cell is saturated, MEG can not cope with  the increased number of requests and tend to drop a large number of users, while MWDG manages to take advantage of the increased multi-user diversity to satisfy a large fraction of the requests.
\par

Fig.~\ref{fig:loadk2} shows the average number of PRBs assigned per user $\eta$  for RG, MEG and MWDG with \emph{uniform power allocation} (UPA) and DPRA. The results for FFR-B are not plotted since for this algorithm the power allocation is fixed and DPRA can not be used. Since the strategy of MEG is to allocate first the users with the worst channel gains, the average number of PRBs per user is relatively high and, as the number of users grows,  is only marginally reduced by DPRA. MWDG, on the other hand, manages to benefit from the increased multi-user diversity and $\eta$ tends to one as $N$ is larger than the number of PRBs. On average, employing DPRA reduces the value of $\eta$  and the cell load by 10\% for MWDG and even more for RG.
\par

Figs.~\ref{fig:dropk3} and  \ref{fig:loadk3} show for $M=3$ the number of dropped users and the  average number of PRBs per user for the various algorithms. The trends exhibited for $M=2$ are confirmed for $M=3$ with MWDG that outperforms all other algorithms. For high cell loads the average number of PRBs per user $\eta$ is practically one when employing DPRA thanks to the algorithm capacity of exploiting  the large diversity of the system.
\par
The differences between the results obtained for $M=2$ and $M=3$ are only marginal. This suggests that for most applications $M=2$ strikes a good balance between efficiency and complexity of the allocation.
\begin{figure}[!htp]
    \centering
    \includegraphics[width=3in]{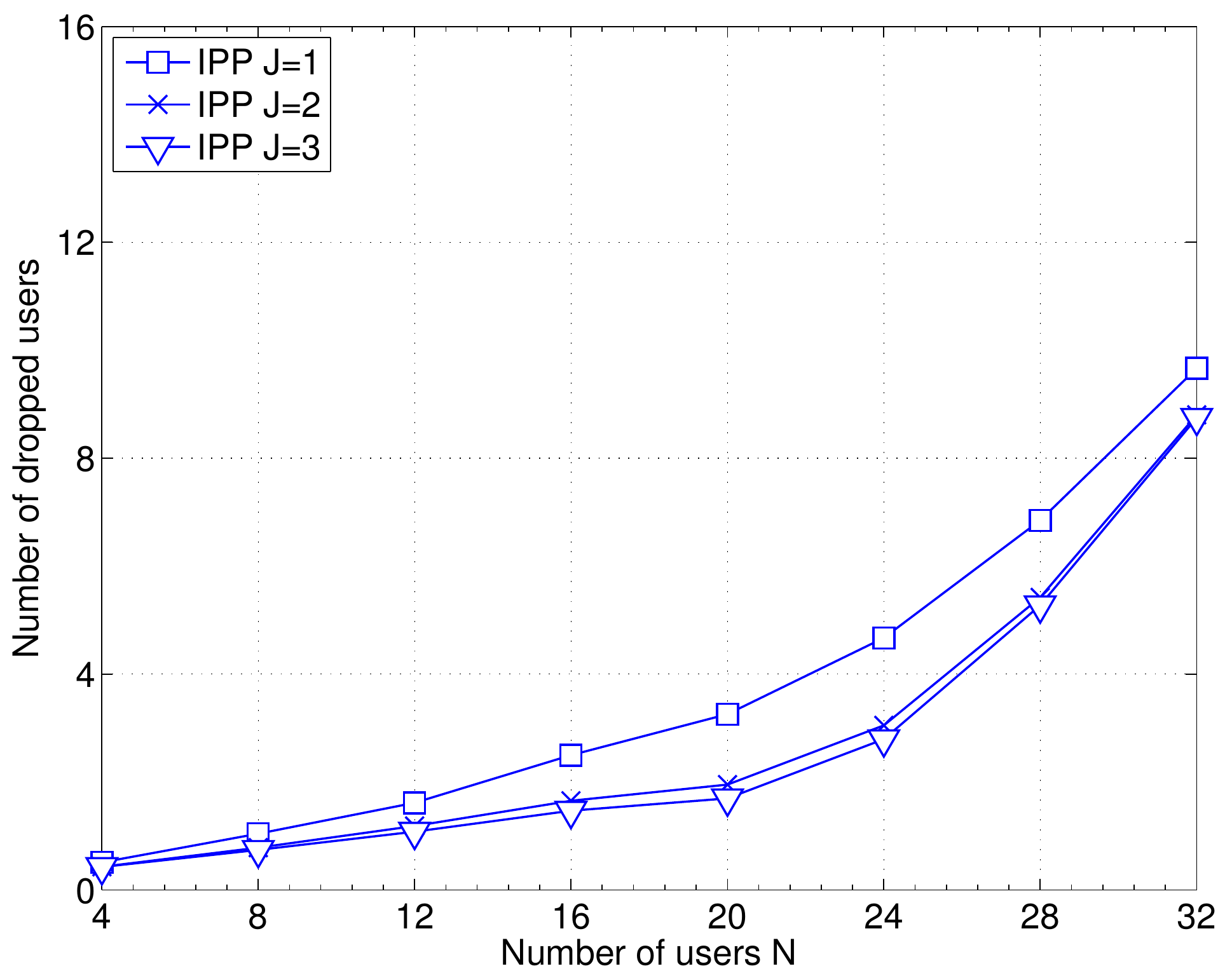}
     \caption{Number of dropped users vs.  number of users when for the IPP allocation, $M=2$}
     \label{fig:Perfite}
\end{figure}
\par
The gains of MWDG in combination with DPRA motivate the implementation of an iterative strategy, which improves the allocation performance by employing the power distribution obtained with DPRA, in place of the  uniform  power distribution, to compute the PRB gains in \eqref{eqs1}. Thus, each iteration of the \emph{iterative PRBs and power}  (IPP) allocation requires first running the MWDG algorithm and then the DPRA. Fig.~\ref{fig:Perfite} shows the results for IPP allocation obtained for $J=1,2$ and $3$ iterations and $M=2$. The results for $J=1$ are equivalent to those plotted in Fig.~\ref{fig:dropk3}. The gains after $J= 2$ iterations are minimal so that it seems that IPP with two iterations represents a good compromise between complexity and performance with very low number of dropped users also for high cell loads.

\section{Conclusions}
\label{secCon}
In this paper, we have  proposed a practical allocation algorithm designed to allocate radio resources in future generation OFDMA cellular systems. The allocation has been formulated with a double objective: either minimizing the cell load when all users can be served or otherwise maximizing the number of served users. For the sake of practical implementation allocation has been decomposed into two sub-problems: radio blocks allocation under deterministic power distribution and power assignment. We have proved that the PRB allocation problem is a minimum weighted independent set (MWIS) problem and proposed a low complexity  heuristic solution to solve it, the minimum weighted degree greedy (MWDG) algorithm. The performance of the MWDG algorithm has been bounded by means of an approximation ratio, which depends on the total number of PRBs and the maximal PRBs number allocated to each user. The power allocation problem has been solved with the objective of further reducing the number of occupied radio resources.
The main characteristic of the proposed scheme is that its complexity is shared between terminals and base station and the amount of control information to be exchanged is extremely limited compared to other schemes.
Simulations have been carried out under constant bit rate traffic model and the results have been compared with other allocation schemes of similar complexity. MWDG has excellent performance and outperforms all other techniques.

\section*{Appendix A: a bound for $W_{MWIS}\left(\tilde{ H}^{(j)} \right)$}
\par
Let $G^{(j)}\left(V^{(j)},E^{(j)},W\right)$ be the graph at the $j$th iteration of the  MWDG algorithm  vertex and $m^{(j)}$ be the vertex found at iteration $j$ of the MWDG algorithm. Moreover, let the user $u$ be associated with vertex $m^{(j)}$ and $\mathcal{M}^{(j)} \subseteq V^{(j)}$ be the set that contains the vertex $m^{(j)}$ and all the vertices  adjacent. The weighted degree of  vertex $m^{(j)}$ is computed as
\begin{equation}
\label{eqAppendix1}
{d_w^{G^{(j)}}}\left( m^{(j)} \right) =\frac{W^{G^{(j)}}\left(\mathcal{C}_{u}\right)-W\left( m^{(j)}\right)+{W}^{G^{(j)}}\left(\mathcal{M}^{(j)}\backslash \mathcal{C}_{u}\right)}{W\left( m^{(j)}\right)}
\end{equation}
where ${W}^{G^{(j)}}\left(\mathcal{M}^{(j)}\backslash \mathcal{C}_{u}\right)$ denotes the sum weight of neighbor vertices of $m^{(j)}$ in $G^{(j)}$ which do not belong to $\mathcal{C}_{u}\subseteq V^{(j)}$ and $W^{G^{(j)}}\left(\mathcal{C}_{u}\right)$ denotes the sum weight of clique $\mathcal{C}_{u}$.

The weighted degree of a  vertex $v$ associated to user $ k \neq u$ such that $ v\in\mathcal{P}_{k}(m^{(j)})$ in $G^{(j)}$,  is computed as as
\begin{equation}
\label{eqAppendix2}
{d_w^{G^{(j)}}}\left( v \right) =\frac{W^{G^{(j)}}\left(\mathcal{C}_{k}\right)-W\left( v\right)+{W}^{G^{(j)}}\left(\mathcal{M}^{(v)}\backslash \mathcal{C}_{k}\right)}{W\left( v\right)}
\end{equation}
where, in analogy with the notation employed above, $\mathcal{M}^{(v)}\subseteq V^{(j)}$ indicates the the set that contains the vertex $v_{k}$ and all the vertices  adjacent. Let $Y\left(v\right)$ be the associated PRBs allocation set of vertex $v$. If $|Y\left(v\right)|=1$, then  it is
\begin{equation}
\label{eqAppendix3}
Y\left(v\right)\subseteq Y\left(m^{(j)}\right)
\end{equation}
\begin{equation}
\label{eqAppendix4}
\left\{\mathcal{M}^{(v)}\backslash \mathcal{C}_{k}\right\}\subseteq \left\{\mathcal{M}^{(j)}\backslash \mathcal{C}_{u}\right\}
\end{equation}
If $\left\{\mathcal{M}^{(v_k)}\backslash v_k\right\}\subseteq \left\{\mathcal{M}^{(j)}\backslash m^{(j)}\right\}$, then ${d_w^{G^{(j)}}}\left( v_k \right)<{d_w^{G^{(j)}}}\left( m^{(j)} \right)$ which contradicts the fact that \emph{the weighted degree of  vertex $m^{(j)}$ is minimal}. Hence, $\left\{\mathcal{M}^{(v_k)}\backslash v_k\right\}\not\subset \left\{\mathcal{M}^{(j)}\backslash m^{(j)}\right\}$. According to \eqref{eqAppendix4}, we can obtain $\mathcal{C}_{k}\not\subset \left\{\mathcal{M}^{(j)}\backslash m^{(j)}\right\}$. Then, there exists a vertex in $\mathcal{C}_{k}$ that is not adjacent to $m^{(j)}$. When we select $m^{(j)}$ into the MWIS, the vertices in $\mathcal{M}^{(j)}$ will be removed from graph $G^{(j)}$. However, user $k$ who constructs clique $\mathcal{C}_{k}$ will not be dropped since there exists a vertex in $\mathcal{C}_{k}$ that does not belong to $\mathcal{M}^{(j)}$.

Moreover, if $|Y\left(v\right)|>1$, $Y\left(v\right)\not\subset Y\left(m^{(j)}\right)$ and $\mathcal{C}_{k}\subseteq \left\{\mathcal{M}^{(j)}\backslash m^{(j)}\right\}$ may hold. When we select $m^{(j)}$ into the MWIS, the vertices in $\mathcal{M}^{(j)}$ will be removed including $\mathcal{C}_{k}$ and user $k$ will be dropped. Since the weight of $m^{(j)}$ is $W\left( m^{(j)}\right)$, there will be no more than $|\pi|-W\left( m^{(j)}\right)$ vertices of MWIS in $\left\{\mathcal{M}^{(j)}\backslash \mathcal{C}_{u}\right\}$. Besides, since vertex $m^{(j)}$ has been selected into MWIS, there will be no more than $|\pi|-W\left( m^{(j)}\right)-1$ vertices which are missed for MWIS. Hence, $W_{MWIS}\left(\tilde{ H}^{(j)} \right)$ can be bounded as \eqref{eq2-10}.

Here, vertices $v_n\in\mathcal{C}_{u}$ with $Y\left(v_n\right) \bigcap Y\left(m^{(j)}\right)=\emptyset$ are not considered. The weighted-degree of any  vertex $v_n\in\mathcal{C}_u$ that does not share PRBs with vertex $m^{(j)}$ in  graph $G^{(j)}$ is larger than ${d_w^{G^{(j)}}}\left( m^{(j)} \right)$ ,
\begin{eqnarray}
\label{eqap-1}
&{d_w^{G^{(j)}}}\left( v_n \right) = \frac{ W^{G^{(j)}}\left( \mathcal{C}_u\right)-W(v_n)+{W}^{G^{(j)}}\left(\mathcal{M}^{(v_n)}\backslash \mathcal{C}_{u}\right)}{W(v_n)} \nonumber\\
& \geq \frac{W^{G^{(j)}}\left(\mathcal{C}_{u}\right)-W\left( m^{(j)}\right)+{W}^{G^{(j)}}\left(\mathcal{M}^{(j)}\backslash \mathcal{C}_{u}\right)}{W\left( m^{(j)}\right)}=d_w(m^{(j)}).
\end{eqnarray}
If $|Y(v_n)|<|Y(m^{(j)})|$, then:
\begin{eqnarray}
\label{eqap-2}
W(v_n)> W(m^{(j)})\;,
\end{eqnarray}
and according to equation \eqref{eqap-1}, it is
\begin{eqnarray}
\label{eqap-3}
{W}^{G^{(j)}}\left(\mathcal{M}^{(v_n)}\backslash \mathcal{C}_{u}\right)>{W}^{G^{(j)}}\left(\mathcal{M}^{(j)}\backslash \mathcal{C}_{u}\right)\;.
\end{eqnarray}
According to equation \eqref{eqap-3}, $\left\{\mathcal{M}^{(v_n)}\backslash \mathcal{C}_{u}\right\}\not\subset  \left\{\mathcal{M}^{(j)}\backslash \mathcal{C}_{u}\right\}$
and there  exists more than one vertex $v\in \left\{\mathcal{M}^{(v_n)}\backslash \mathcal{C}_{u}\right\}\;\&\;  v\notin \left\{\mathcal{M}^{(j)}\backslash \mathcal{C}_{u}\right\}$. Hence, the selecting of $m^{(j)}$ doesn't influence the cardinality of MWIS and vertex $v_n$ can be neglected.

Besides, if $|Y(v_n)|\geq|Y(m^{(j)})|$, then:
\begin{eqnarray}
\label{eq2-12bis}
W(v_n)\leq W(m^{(j)})\;.
\end{eqnarray}

Hence, ${W}^{G^{(j)}}\left(\mathcal{M}^{(v_n)}\backslash \mathcal{C}_{u}\right)<{W}^{G^{(j)}}\left(\mathcal{M}^{(j)}\backslash \mathcal{C}_{u}\right)$ and $\left\{\mathcal{M}^{(v_n)}\backslash \mathcal{C}_{u}\right\}\subseteq  \left\{\mathcal{M}^{(j)}\backslash \mathcal{C}_{u}\right\}$ may be satisfied. When vertex $m^{(j)}$ is selected into MWIS, vertices belonging to $\left\{\mathcal{M}^{(v_n)}\backslash \mathcal{C}_{u}\right\}$ will be removed as well as vertex $v_n$. However, since $\left\{\mathcal{M}^{(v_n)}\backslash \mathcal{C}_{u}\right\}\subseteq  \left\{\mathcal{M}^{(j)}\backslash \mathcal{C}_{u}\right\}$ and $ |\pi| - {M} \le  W\left(v_n\right) \le |\pi| - 1$,  there will be no more than $M$ vertices of MWIS in $\mathcal{M}^{(v_n)}\cup \mathcal{M}^{(j)}$. Hence, $W_{MWIS}\left(\tilde{ H}^{(j)} \right)$ can be bounded as \eqref{eq2-10} and vertex $v_n$ can be neglected.
\section*{Acknowledgment}
This work was supported by  Multi-cell and Multi-user Interference Mitigation in IMT-Advanced, Supported by MIIT of China, No.2010ZX03003-002. (Jan 2010 $\sim$ Dec 2012)
and MAC in Asymmetrical Wireless Network, Supported by Education Bureau of Anhui Province, No.KJ2010A333. (Jan 2010 $\sim$ Dec 2011)
\bibliographystyle{IEEEtran}

\bibliography{IEEEabrv,bib}
\end{document}